\documentclass[aps,prl,amsmath,twocolumn,amssymb,superscriptaddress]{revtex4}
\pdfoutput=1
\usepackage{graphicx}
\usepackage{epstopdf}
\usepackage{color}
\usepackage{cases,array}

\begin{document}
\title{Mutual Phase Locking of Very Nonidentical Spin Torque Nanooscillators via Spin Wave Interaction}

\author{A.R. Safin}
\affiliation{National Research University "Moscow Power Engineering Institute", 112250 Moscow, Russia}

\author{N.N. Udalov}
\affiliation{National Research University "Moscow Power Engineering Institute", 112250 Moscow, Russia}

\author{M.V. Kapranov}
\affiliation{National Research University "Moscow Power Engineering Institute", 112250 Moscow, Russia}

\begin{abstract}
In this paper the mutual phase locking theory of very nonidentical spin-torque nanooscillators, which is based on the Slavin-Tiberkevich model, considering the theory of nonlinear oscillations, is developed. Using generalized Adler equation we calculate phase-locking region of the system with spin-wave coupling in the parameter plane - distance between nanocontacts and radii difference. We describe trajectories of such a system in the phase space and show the effect of a broadband synchronization. We introduce a generalization of this approach to the ensembles of spin-torque nanooscillators.

\end{abstract}

\maketitle

\section{I Introduction}
The spin-transfer torque effect in the magnetic multilayers was theoretically predicted by J. Slonczewski [1] and L. Berger [2] in 1996. This effect will give a chance to implement new methods of microwave generation in a nanoscale. The scientists previously showed (for more details see perfect review [3]), that electrical current, which is passing through a magnetic multilayered structure becomes spin-polarized and, if the current density is high (near then 107-108 A/cm2), the spin-polarized current can transfer spin angular momentum among this magnetic layers, which leads to the microwave generation of such a structure.

Microwave spin transfer torque - based autooscillators, which are called spin torque nanooscillators (STNO), are very attractive for potential applications. They are highly tunable by bias current and magnetic field, they are the smallest oscillators that have ever been developed (more than 50 times smaller than a standard autooscillators), and can be biased at low currents and voltages (less than 1.0 V). The main drawback of STNOs is its very weak output microwave power (less than 1 µW for a simple structure of layers). This negative effect creates some difficulties for the development of novel nanosized devices based on STNOs.

A suggested solution of this problem is to synchronize several STNOs and to coherently summarize an output power from each device. Latter effect gives the way to coherent addition of microwave signals by partial nanocontacts. The coupling between the STNOs can be caused by different physical mechanisms. One of the first of them is the nonlocal mechanism (not dependent on a distance between contacts) coupling through the common microwave component of the bias current (for details see [4-7]). Using this mechanism of the mutual phase locking it is rather difficult to achieve impedance matching of each STNO as a part of an ensemble, and it is really possible to achieve synchronization of only a several tens of oscillators that may not be enough for practical realization. Another [3] physical mechanism of mutual coupling of STNOs is the local mechanism which strongly depends on the distance between nanocontacts. Compared with common bias current phase locking mechanism, this mechanism can give a much more size compactification effect in the creation of a large ensembles. One of the simplest local coupling mechanisms of the STNOs is the dipole-dipole mechanism, which takes place both for the nanocontacts and nanopillars. One of the most perspective type of STNO's local coupling is the coupling through the radiation of the propagating spin waves for nanocontacts. In the nanocontact type of STNOs geometry at a distance from 150 to 500 nm [3] interaction via spin waves is much stronger than the interaction through the dipole-dipole interaction. This phenomenon of mutual phase locking for a two STNOs via spin waves has experimentally been observed in [8-10]. Theoretical models of two point contacts STNOs, based on the classical quasi-Hamiltonian formalism for the spin waves were presented by A. Slavin, V. Tiberkevich in [11] and S.M. Rezende, et al [12]. The theoretical analysis taking into account of time delay between spin waves was presented in [3]. In real experiments, it is not possible to create the ensembles of STNO with fully identical parameters (especially diameters). Therefore, it is very important to analyze nonlinear dynamics of the autooscillators with nonidentical parameters using general oscillation theory.

In a classical literature of oscillation theory (see for example [13-15]) dynamics of coupled van der Pole oscillators and van der Pole-Duffing oscillators are regarded as basic models of phase locking effect. Usually, the pay a special attention to the identical or small nonidentical cases. However, as it recently has been shown [16] for the coupled van der Pole oscillators, this classical problem is more complicated and interesting. The corresponding equations for small variable amplitudes of spin waves for STNOs are more difficult than classical van der Pole equations (nonisochronous, phase delay and strongly nonidentical regenerative components).
In [17] the phase locking system of two coupled vortex spin-torque nanooscillators was investigated. They showed, that for relatively small diameter differences the synchronization dynamics can be described qualitatively using Adler equation. However, when the diameters difference increases signi?cantly, the system becomes strongly "non-Adlerian". In a large number of works by "Adlerian dynamics" is understood the system, which can be described by the following equation [13-16]:

\begin{equation}
\frac{\mathrm{d}\Psi}{\mathrm{d}t} = \Delta \omega - A \cdot F(\Psi), \label{eq:commonAdler}
\end{equation}
where $\Delta \omega$  is the initial frequency difference between oscillators,  $A$ is the greatest phase deviation parameter, and $F(\Psi)=\sin(\Psi)$ . In this case the main phase locking area (main Arnold tongue) on the parameter plane $(A,\Delta \omega)$  is the area bounded by straight lines emanating from the point  , which is the theoretical idealization. If the function   is more complicated, the phase locking area becomes much more complex, and the equation (1) in the literature is called "generalized phase equation" or "generalized Adler equation". The method of obtaining the phase equation in [17] with   is based on the fact that the radii of the oscillators were equal to stationary uncoupled radii. This approach is used in a large number of papers on the theory of autooscillators (see for more details [13-15]). For the large ensembles, this approach leads to a Kuramoto [3] model. However, a large radii difference between coupled STNOs leads to the fact that stationary amplitudes depend on the phase difference and to the more difficult function $F(\Psi)$   in (1) and more complex form of phase locking area. Therefore, to describe the phase dynamics of two nonidentical STNO it is necessary to use dependence of a stationary amplitude on a phase difference between oscillators.

In this paper the mutual phase locking theory of very nonidentical spin-torque nanooscillators, which is based on the Slavin-Tiberkevich model (see [3,11,12]), considering the theory of nonlinear oscillations, is developed. The outline of the paper is following.

After an introduction, which is given in the Section I, In the Section II according to [3] we give an analytical model of two coupled STNOs. In the Section III we consider generalized Adler-like equation and calculate phase-locking region of the system with spin-wave coupling in the parameter plane - distance between nanocontacts and relative radii difference. In the Section IV, we consider trajectories of such a system in the phase space (amplitudes and phase difference) and show the effect of a broadband synchronization. We discuss basic bifurcations in the system of coupled STNOs. In the Section V, we present obtaining of generalized Adler-like equations to the ensembles STNOs. Conclusions are represented in the Section VI.

\section{II ANALYTICAL MODEL OF TWO COUPLED SPIN-TORQUE NANOOSCILLATORS}

Our physical model is based on two magnetic nanocontacts of radii $R_{C1,2}$ formed in the common free ferromagnetic layer with thickness $d$  and saturation magnetization $M_0$  , separated by the distance $\rho_0$. For correct research the dynamics of two coupled STNO it is necessary to use system of two coupled Landau-Lifshitz-Gilbert equations. In general, the theoretical analysis of such a system is rather difficult. Therefore, in [11,12] the mathematical model of two coupled STNO as a system of coupled equations for the spin wave complex amplitudes was received. Thus the phase locking of two nonidentical STNO on the single frequency is naturally supposed. We will consider this situation. That is non-identity between STNO such as they are capable to phase lock at a fundamental frequency. We will find the corresponding phase locking area and its border. In order to find the phase locking area boundaries on the other harmonics (for example on the fractional frequencies) it is necessary to obtain the truncated equations on appropriate frequencies that goes outside of this work. The mathematical model (Slavin-Tiberkevich model) of such a system for two-component vector $\mathbf{C} = [c_1,c_2]^T$  of the complex precession amplitudes $c_{1,2}(t)$  in two nanocontacts can be described like this:

\begin{equation}
\frac{\mathrm{d}\mathbf{C}}{\mathrm{d}t} + i\Omega\cdot\mathbf{C} + \Delta \mathbf{C}=0, \label{eq:commonAdler}
\end{equation}
where $ \Omega = \begin{pmatrix} \omega_1(\vert c_1 \vert ^2) & -\Omega_1\sin \beta_1 \\ -\Omega_2\sin \beta_2 & \omega_2(\vert c_2 \vert ^2) \end{pmatrix} $ - matrix of frequencies and $\Delta = \begin{pmatrix} \Gamma_{eff1}(\vert c_1 \vert ^2) & -\Omega_1\cos \beta_1 \\ -\Omega_2\cos \beta_2 & \Gamma_{eff2}(\vert c_2 \vert ^2) \end{pmatrix}$ - matrix of regenerability (effective damping) of this system, $\omega_j(\vert c_j \vert ^2)=\omega_0 + D \cdot k_{Cj}^2 + N\vert c_j \vert ^2$ - partial frequency of j-th STNO, $\Gamma_{eff,j}(\vert c_j \vert ^2)=-\Gamma_G(\zeta_j+q)(a_j-\vert c_j \vert ^2)$ - effective damping of individual STNO. Here $\omega_0=\sqrt{\omega_H(\omega_H+\omega_M\cos^2\theta_{int})}$ - frequency of linear ferromagnetic resonance, $\omega_H=\gamma H$, $\omega_M=4\pi\gamma M_0$, $\gamma = 2.8 MHz/Oe$ - is the gyromagnetic ratio, $D=\omega_M \lambda_{ex}^2(\partial\omega_0/\partial\omega_H)$ is the dispersion coefficient of spin waves (here $\lambda_{ex}=5nm$ - exchange length), $k_{C,j}=1.2/R_{C,j}$ - wave number, $N=\frac{\omega_H\omega_M}{\omega_0}(\frac{3\omega_H^2\sin^2\theta_{int}}{\omega_0^2}-1)$ - nonlinear frequency coefficient (nonisochronism parameter), $\Gamma_G=\alpha_G\omega_0(\partial\omega_0/\partial\omega_H)$ is the linear Gilber damping, $\alpha_G$ is the linear damping coefficient, $\zeta_j=I_j/I_{th,j}$ - - supercriticality parameter of j-th STNO, $I_j$ is the bias current (in mA) which is passed through j-th STNO, $I_{th,j}=\Gamma_G/\sigma_j$ is the threshold current, $\sigma_j=(\epsilon_j g \mu_B)/(2e M_0 V_{eff,j})$, $\epsilon_j$ is the spin-polarization efficiency in the j-th nanocontact, $g$ is the Lande factor, $e$ is the modulus of the electron charge, $V_{eff,j}$ is the effective volume of the j-th contact, $q$ is the nonlinear damping coefficient, $a_j=(\zeta_j-1)/(\zeta_j+q)$ is the partial stationary spin-wave power of j-th STNO. As was shown in [3,11] the effective $V_{eff,j}$ of the j-th nanocontact can be expressed in the following form:

\begin{equation}
V_{eff,j}=(1+1.86\lambda_{ex}^2 R_{Cj}^2 \frac{\omega_M}{\Gamma_G})\cdot V_j, \label{eq:Volume}
\end{equation}
where $V_j=\pi R_{Cj}^2 d$. In Eqs. (2)-(3), $H_{int}$ and $\theta_{int}$ are the internal bias magnetic field magnitude and out-of-plane angle respectively, which are connected with the external magnetic field magnitude $H_{ext}$ and angle $\theta_{ext}$ by following expressions:
\begin{subequations}
\begin{align}\label{Magn}
H_{ext}\cos\theta_{ext} = H_{int} \cos \theta_{int}; \\ 
H_{ext}\sin\theta_{ext} = (H_{int}+4\pi M_0)\sin\theta_{int}. 
\end{align}
\end{subequations}
In equation (2) the coupling frequencies $\Omega_{1,2}$ and coupling phases $\beta_{1,2}$ are equal to
\begin{gather}
\Omega_{1,2} = 0.65 \cdot \Gamma_G \sqrt{\frac{R_{C1,2}}{\rho_0}} \exp ( -\Gamma_G \rho_0/ v_{gr1,2} ), \\
\beta_{1,2} = \rho_0 \cdot k_{C1,2}, 
\end{gather}
where $v_{gr1,2}=1.7\omega_M \lambda_{ex}^2 / R_{C1,2}$  is the group velocity of the excited propagating spin wave mode.

Let's rewrite system (2) under amplitudes $U_{1,2}$  and phase difference $\psi = \phi_1 - \phi_2 + 0.5(\beta_1-\beta_2)$  of spin waves using expression $c_{1,2}(t)=U_{1,2}(t)\exp(-i \cdot \phi_{1,2}(t))$  in the following form
\begin{subequations}
\begin{align}\label{Trunc}
\frac{dU_1}{dt}=-U_1\cdot \Gamma_{eff1}(U_1)+\Omega_1 U_2 \cos(\psi+\delta \beta); \\
\frac{dU_2}{dt}=-U_2\cdot \Gamma_{eff2}(U_2)+\Omega_2 U_1 \cos(\psi-\delta \beta); \\
\frac{d\psi}{dt}=\Delta\omega-(\Omega_1\frac{U_2}{U_1}\sin(\psi+\delta\beta)+\Omega_2\frac{U_1}{U_2}\sin(\psi-\delta\beta)).
\end{align}
\end{subequations}
where $\delta\beta=0.5(\beta_1+\beta_2)$  is the average phase displacement between complex amplitudes and $\Delta\omega=D(k_{C1}^2-k_{C2}^2)+N(U_1^2-U_2^2)$ - frequency difference between complex amplitudes.

\section{III PHASE DYNAMICS: GENERALIZED ADLER EQUATION}

Let's assume that   that nonidentity, the nonisochronous parameter and the coupling have the same order of smallness. Equations for the amplitudes of uncoupled STNOs can be written as follows $dU_{1,2}/dt=-U_{1,2}\Gamma_{eff1,2}(U_{1,2}^2)$ . They have a stable stationary solution corresponding to orbits $U_{1,2}^0 = \sqrt a_{1,2}$ . Now let's consider perturbed solutions with the following amplitudes $U_{1,2}=U_{1,2}^0 + \xi_{1,2}$ , where $\xi_{1,2}$ are the perturbation terms. After substituting these expressions into the (7a) and (7b) and after ignoring high order terms (for more details see [16]), it is possible to find stationary perturbation terms $\xi_{1,2}^0$ in the following form:

\begin{equation}
\xi_{1,2}^0=\frac{\Omega_{1,2}U_{2,1}^0}{2U_{1,2}^0\Gamma_G(\zeta_{1,2}+q)}\cos(\psi\pm \delta\beta). \label{eq:xi}
\end{equation}
In the next step let's substitute expression $U_{1,2}=U_{1,2}^0 + \xi_{1,2}$ into the (7c) we will consider further "truncated" phase equation (generalized Adler-like equation) in the form (1) with complicated function $f(\psi)$. In this equation we have (using expression $1/(1+x)\approx 1-x$)

\begin{gather}
\Delta\omega=D(k_{C1}^2-k_{C2}^2)+N(a_1-a_2)+N(A_1\cdot \\
\cdot \cos(\psi+\delta\beta)-A_2\cos(\psi-\delta\beta)); \nonumber \\
\frac{U_{1,2}}{U_{2,1}} = \frac{U_{1,2}^0}{U_{2,1}^0} +\frac{1}{2U_1^0U_2^0}(A_{1,2}\cdot \\
\cdot \cos(\psi \pm \delta\beta)-A_{2,1}\frac{a_{1,2}}{a_{2,1}}\cos(\psi \mp \delta\beta)),\nonumber
\end{gather}
where $A_{1,2}=\frac{\Omega_{1,2}U_{2,1}^0}{\Gamma_G (\zeta_{1,2}+q)}$  are the constants. So final Adler-like equation has the form (7c) with (9), (10).  It is convenient to write final a Adler-like equation in the following form:

\begin{equation}
\frac{d\psi}{dt}=-\frac{\partial W}{\partial \psi}. \label{eq:xi}
\end{equation}
where $W(\psi)$  is the effective potential. Minimum of function $W(\psi)$  gives a rise to the stable synchronouse (phase-locked) regime of oscillations, while maximum corresponds to the asynchronous regimes.
Using numerical analysis of a generalized Adler-like equation we calculate phase-locking regions (Arnold tongues) of two coupled system by spin-wave interaction in the parameter plane $(\rho_0,\delta)$, where $\delta=(R_{C2}-R_{C1})/R_{C1}$  is the relative radii difference between nanocontacts. These dependences are presented at Fig.1a. Here $\kappa = 1 - \zeta_2/\zeta_1$  is the constant, which characterizes a ratio between supercriticality parameters of STNOs.

We can see that if we decrease parameter $\kappa$, then phase locking bandwidth of such a system proportionally increases. At the small values of $\kappa$ the bandwidth of Arnold tongue corresponds to non-zero difference $\delta\not= 0$ in the sizes of the nanocontacts. This is a significant moment,  which reveals that in fully identical scheme (with identical radii) in some situations a phase locking regime may not exist. An extreme case, when $\zeta_1=\zeta_2$  (i.e. $\kappa=0$) leads to a very high phase locking bandwidth. But in this case the center of Arnold tongue also does not correspond to the fully identical scheme with $\delta=0$.

The phase-locking bandwidths in terms of radii differences $\delta_0 = (R_{C2,max}-R_{C1})/R_{C1}$ , which depends on the distance between STNOs for different $\kappa$ are presented in Fig.1b. It can be clearly seen, that if we decrease parameter $\kappa$  the maximum value of $\delta_0$  proportionally increases. So, the practical result of this research is following: if technically nonidentity and contacts distance is fixed, then we get a possibility of extension phase-locking bandwidth by varying of currents $I_{1,2}$.

\begin{figure*}[ht!]
\centerline{\includegraphics[width=170mm]{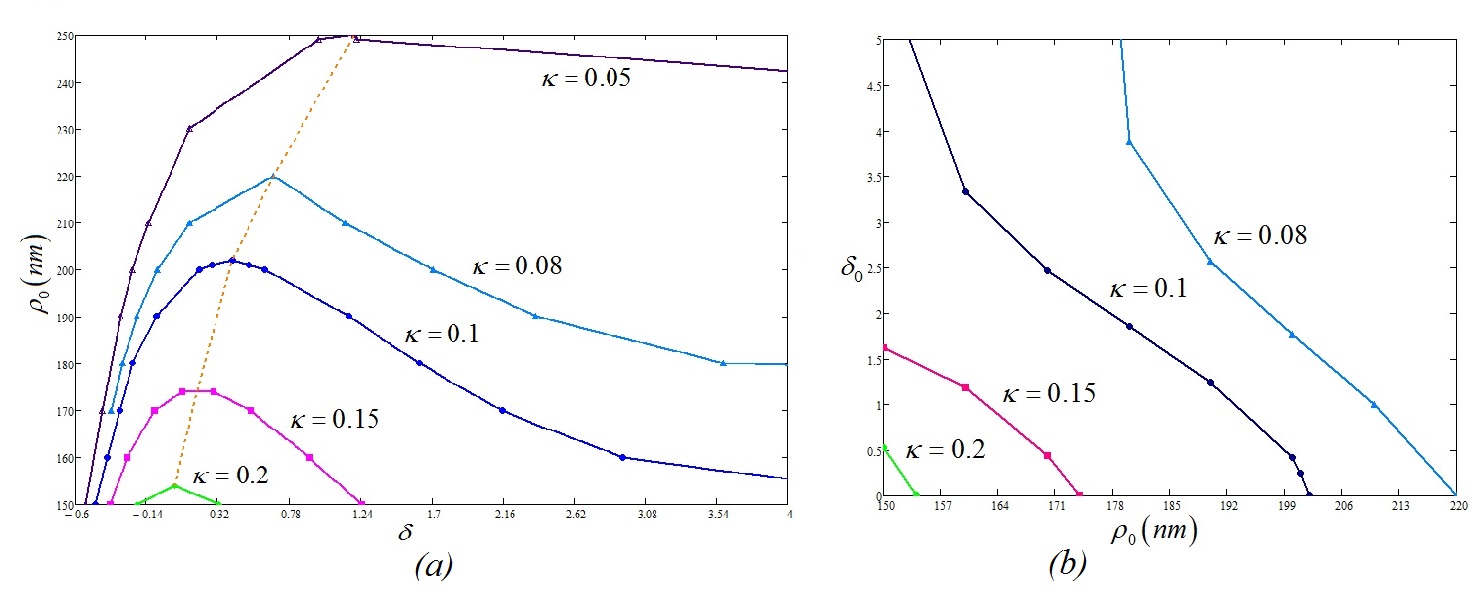}}
\caption{\footnotesize{(Color online) Phase locking regions (Arnold tongues) of the system with spin-wave coupling in the parameter plane ($\rho_0,\delta$) for several values of currents $\kappa=1-\zeta_2/\zeta_1$ (a) and phase-locking bandwidth (b) in terms of radii difference $\delta_0$  which depends on the distance between STNOs $\rho_0$. Parameters of this model: $\theta_{ext}=\pi/2$, $H_{ext}=1T$, $4\pi M_0=8kG$, $\alpha_G=0.01$, $R_{C1}=100nm$, $\epsilon_{1,2}=0.1$, $g=2$. Dotted line on (a) is the center of symmetry of each Arnold tongue.}}
\label{fig:fig2}
\end{figure*}

Analyzes of the generalized Adler equation gives a good agreement with analyzing of an original system (6) but much easier then (6). However, this approach does not give a possibility to analyze a special amplitude dynamics of coupled STNOs near border line of a synchronization zone.

Let's discuss the questions of two coupled STNO effective spectral linewidth of the output signal. In experiments [8-10] it was shown, a significant increase of the output power, an ensemble of mutually phase-locked STNOs has a much smaller generation linewidth than a single oscillator in the same array, which is also a very important advantage of this scheme. We can assume, that in the symmetry center of the corresponding phase locking the spectral linewidth of the output signals is minimal, and near the boundary it's begins to increase. For the quantitative determination of a spectral linewidth of two phase locked STNO it is necessary to insert in (2) an additional noise vector $\Theta = [\xi_1,\xi_2]^T$  and to use standard methods of the statistical theory (see e.g. [3] for a single STNO) that goes beyond of this work and makes a subject of a separate study.

In the next Section we will demonstrate a special nonlinear dynamics (in phase space) of such a scheme using equations (6).

\section{IV	TRAJECROTIES IN PHASE SPACE}

Now, let's  refer to the phase portraits of the coupled STNO system (eq. (6)) in a different situation. Firstly, the phase portrait of this system in the center of phase locking zone is presented in Fig.2a. Here we have one stable node point "1" and one non-stable saddle point "2".

\begin{figure*}[ht!]
\centerline{\includegraphics[width=170mm]{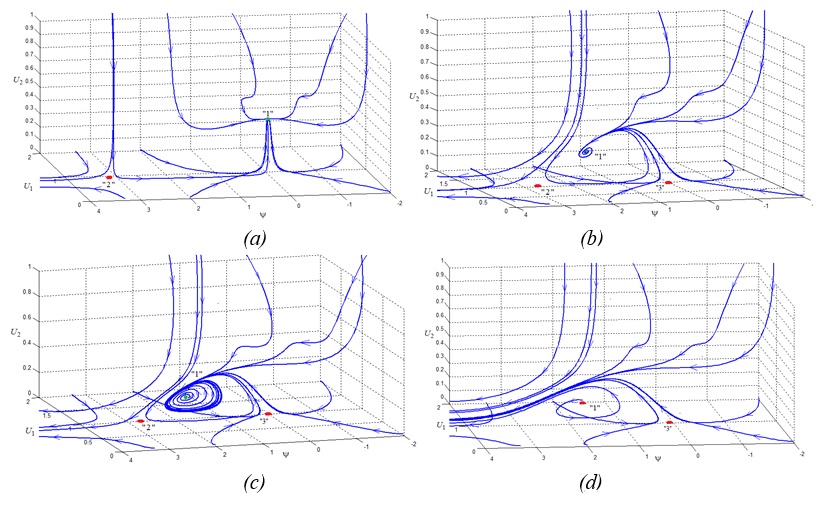}}
\caption{\footnotesize{(Color online) Phase portraits of the coupled STNO system in the center of phase locking zone (a), near the border of synchronization zone, in which stable node turns into to the stable focus (b) and after increasing of a frequency difference limit cycle starts to form (c). After increasing a frequency difference between STNOs this system transfers to asynchronous regime (d).}}
\label{fig:fig2}
\end{figure*}

After increasing of frequency difference between STNOs a stable node turns into the stable focus (Fig.2b) and loses its stability after some value of nonidentically  $\delta_{cr}$. In this case a limit cycle starts to form (Fig.2c). The system in this case is in the broadband regime in which one of oscillators dominates the other STNO. Amplitude of one of oscillators asymptotically decreases. After increasing a frequency difference between STNOs this system transfers to asynchronous regime (see Fig.2d). This broadband regime in the Arnold tongue can be seen at very nonidentical parameters (at high  $\delta$). In the circumstances of real life, this effect does not fit very well to the power increasing, because an amplitude of weaker oscillator is much smaller than amplitude of the other oscillator. Moreover, efficiency of equivalent scheme of this power increasing will not be high. It's more relevant for using scheme of coupled STNOs to use it in a center of a synchronous regime, but it is possible not only for identical case.

\section{V	PHASE DYNAMICS: GENERALIZATION TO THE ENSEMBLES OF OSCILLATORS}

Now, let's consider generalization of this approach to the mutually coupled STNOs  ensemble. In this case model (2) is also applicable to the N-dimensional vector $\mathbf C = [c_1,...,c_N]^T$ , where N is the dimension of a system. Here for a ladder line ensemble of STNOs we have following expressions for the matrices $\Omega$  and $\Delta$:

\begin{gather}
\Omega = \begin{pmatrix} \omega_1 & -\Omega_{1,2}\sin\beta_{12} & 0 & \dots & 0 \\
-\Omega_{21}\sin\beta_{21} & \omega_2 & -\Omega_{23}\sin\beta_{23} & \dots & 0 \\
0 & -\Omega_{32}\sin\beta_{32} & \omega_3 & \dots & 0 \\
\vdots & \vdots & \vdots & \vdots & \vdots \\
0 & 0 & 0 & \dots & \omega_N
  \end{pmatrix};\\
\Delta = \begin{pmatrix} \Gamma_1 & -\Omega_{1,2}\cos\beta_{12} & 0 & \dots & 0 \\
-\Omega_{21}\cos\beta_{21} & \Gamma_2 & -\Omega_{23}\cos\beta_{23} & \dots & 0 \\
0 & -\Omega_{32}\cos\beta_{32} & \Gamma_3 & \dots & 0 \\
\vdots & \vdots & \vdots & \vdots & \vdots \\
0 & 0 & 0 & \dots & \Gamma_N
  \end{pmatrix},
\end{gather}
where $\Omega_{i,j}$, $\beta_{i,j}$ are the coupling frequency and the coupling phases which characterize the influence of the j-th to i-th STNO and $\Gamma_j=\Gamma_{eff,j}$. In this case, we can write truncated equations under amplitudes $U_i$ and phase differences $\psi_j = \phi_j - \phi_{j-1} + 0.5(\beta_{j,j-1}-\beta_{j-1,j})$ in the following form:

\begin{gather}
\frac{dU_i}{dt} = -U_i\Gamma_i+\Omega_{i,i+1}U_{i+1}\cos(\psi_i+\delta\beta_i)+ \\
+ \Omega_{i,i-1}U_{i-1}\cos(\psi_i-\delta\beta_i); \nonumber \\
\frac{d\psi_j}{dt} = \Delta\omega_j - (\Omega_{j,j-1}\frac{U_{j-1}}{U_j}\sin(\psi_j+\delta\beta_j)+\\
+ \Omega_{j-1,j}\frac{U_{j}}{U_{j-1}}\sin(\psi_j-\delta\beta_j))+ \nonumber\\ \Omega_{j,j+1}\frac{U_{j+1}}{U_j}\sin(\psi_{j+1}-\delta\beta_{j+1})+ \nonumber \\ 
+ \Omega_{j-1,j-2}\frac{U_{j-2}}{U_{j-1}}\sin(\psi_{j-1}-\delta\beta_{j-1}). \nonumber
\end{gather}
Here for amplitudes $i=1,\dots,N$ and for phase differences $j=1,\dots,N-1$, and $\Delta\omega_j=\omega_j-\omega_{j-1}$. Now let's consider perturbed solutions with the following amplitudes $U_i=U_i^0+\xi_i$, where $\xi_i$ are the perturbation terms. After substituting these expressions into the (14) and after ignoring of high order terms it is possible to find stationary perturbation terms $\xi_i^0$ in the following form:

\begin{gather}
\xi_i^0=\frac{\Omega_{i,i+1}U_{i+1}^0}{2U_i^0\Gamma_G}(\zeta_i+q)\cos(\psi_i+\delta\beta_i)+ \\
+ \frac{\Omega_{i,i-1}U_{i-1}^0}{2U_i^0\Gamma_G}(\zeta_i+q)\cos(\psi_i-\delta\beta_i). \nonumber
\end{gather}

Now, after substituting expression $U_i=U_i^0+\xi_i^0$ into the (15), we will consider further "truncated" phase equation (generalized Adler-like equation). In a general case this equation has a complicated form, and one of the interesting future perspectives to focus on is to analyze dynamics in a low-dimensional ensembles of STNOs (for N=3 and N=4) and to consider limited case than $N\rightarrow\infty$. This approach differs to the Kuramoto model (see [3]) because it allows an easy change of amplitude with the respect to the coupling between STNOs.

\section{VI	CONCLUSION}

In this paper, we investigate the nonlinear dynamics of a two nonidentical coupled non-contact STNO connected via spin waves. Approximate analysis of this system using a generalized phase equation (Adler like) is proposed. It differs from the standard Adler approach, which is used in most of the literature, that for significantly nonidentical oscillators it is necessary to consider the dependence of the stationary amplitude on a phase difference instead of using uncoupled STNO amplitudes. Using this approach it is obtained phase locking areas by using methods of the nonlinear oscillation theory.

In the paper we have shown that in practice identical scheme of coupled STNOs is not relevant for all values of the parameters. The introduced approach (based on the generalized Adler equation) can give the following result. In the circumstances of real life, this effect does not fit very well to the power increasing, because an amplitude of weaker oscillator is much smaller than amplitude of the other oscillator. Moreover, efficiency of equivalent scheme of this power increasing will not be high. It's more relevant for using scheme of coupled STNOs to use it in a center of a synchronous regime, but it is possible not only for identical case. Furthermore, we displayed that this approach could be applicable to the N-dimensional case.

Our approach allows one to draw a statistical generalization of such a system with the presence of noise, which will determine the spectral linewidth of the output oscillation of two coupled STNO. This task makes a subject of further research.

The obtained dependences and method of determining the phase locking zone give possibility to select the optimum distance between STNO taking into account their essential non-identity.

Authors acknowledge financial support by Russian Foundation for basic research (Grant No. 10-02-01403) and Russian Federal Program "Scientific and scientific-pedagogical Staffs of Innovative Russia" for 2009-2013 (Grant No. 14.B37.21.1211, 14.132.21.1665).

\end{document}